# The Promotion Wall: Efficiency–Equity Trade-offs of Direct Promotion Regimes in Engineering Education

Hugo Roger Paz
PhD Professor and Researcher Faculty of Exact Sciences and Technology National University of Tucumán
Email: hpaz@herrera.unt.edu.ar
ORCID: https://orcid.org/0000-0003-1237-7983

**ABSTRACT**

Progression and assessment rules are often treated as administrative details, yet they fundamentally shape who is allowed to remain in higher education, and on what terms. This article uses a calibrated agent-based model to examine how alternative progression regimes reconfigure dropout, time-to-degree, equity and students' psychological experience in a long, tightly sequenced engineering programme. Building on a leakage-aware longitudinal dataset of 1,343 students and a Kaplan–Meier survival analysis of time-to-dropout, we simulate three policy scenarios: (A) a historical "regularity + finals" regime, where students accumulate exam debt; (B) a direct-promotion regime that removes regularity and finals but requires full course completion each term; and (C) a direct-promotion regime complemented by a capacity-limited remedial "safety net" for marginal failures in bottleneck courses.

The model is empirically calibrated to reproduce the observed dropout curve under Scenario A and then used to explore counterfactuals. Results show that direct promotion creates a "promotion wall": attrition becomes sharply front-loaded in the first two years, overall dropout rises, and equity gaps between low- and high-resilience students widen, even as exam debt disappears. The safety-net scenario partially dismantles this wall: it reduces dropout and equity gaps relative to pure direct promotion and yields the lowest final stress levels, at the cost of additional, targeted teaching capacity.

These findings position progression rules as central objects of assessment policy rather than neutral background. The article argues that claims of improved efficiency are incomplete unless they are evaluated jointly with inclusion, equity and students' psychological wellbeing, and it illustrates how simulation-based decision support can help institutions rehearse assessment reforms before implementing them.

## 1. INTRODUCTION

Universities across the world have struggled for decades with high student dropout, particularly in science, technology, engineering, and mathematics (STEM) programmes. Engineering degrees tend to concentrate academic difficulty, heavy workloads, and rigid curricular structures into the very years in which students are still negotiating their academic identities and living conditions (OECD, 2019; UNESCO-IESALC, 2020). In many Latin American countries, completion rates in engineering remain stubbornly low, with a large fraction of students leaving without a degree, often after several years of investment of time, money, and institutional resources (Paz, 2022; Tinto, 2012). Against that backdrop, institutional leaders increasingly experiment with regulatory reforms—changing assessment regimes, progression rules, or "regularity" requirements—in the hope of improving efficiency and "cleaning up" cohorts (Herd & Moynihan, 2019).

The case studied in this article fits squarely within this broader pattern. In a Civil Engineering programme at a public university, administrative records over fifteen cohorts show that only a small minority of entrants eventually graduate, while a substantial majority either drop out or remain trapped in protracted, debt-laden trajectories with many pending examinations (Paz, 2022). A survival-analysis study of 1,343 students without equivalences showed a median time to dropout of around four and a half years and an eventual dropout proportion above half of the cohort, with a particularly high concentration of exits in the first year (Paz, 2022). This quantitative picture makes visible an uncomfortable reality: the system does not simply "filter" at the start, but gradually leaks students out over many years, producing what has been described as a "leakage" problem rather than a clean selection mechanism (Paz, 2025b).

Until recently, this programme operated under a "regularity + finals" regime. Students would attend and complete courses, obtain a status of "regular" if they passed the continuous assessment, and then face a separate final examination at a later date. In practice, this created an accumulation of "finals debt": students could move forward in the curriculum while carrying unresolved obligations behind them. A previous agent-based simulation study, the "Regularity Trap" experiment, showed how this design creates a structural time-to-live mechanism: once the stock of overdue finals grows beyond a tolerable threshold, the probability of dropout rises sharply, even if the student remains technically enrolled (Paz, 2025a). From an institutional viewpoint, this regime generates an opaque stock of unfinished business that ties up resources and obscures the true state of progression.

In response to these concerns, the institution adopted a very different rule: a promotion-only regime, in which students must fully pass each course—including the summative assessment—to progress, with no lingering finals debt. At first glance, such a reform appears to address an obvious inefficiency. If students are required to complete each course before moving on, they will either progress cleanly or fail early, reducing the long tail of administrative "zombies" who appear in statistics but have effectively disengaged (Herd & Moynihan, 2019). Yet this apparent efficiency gain raises a critical concern: does the new regime simply transform slow "leakage" into a sharp "promotion wall" in the first year, where weaker or more vulnerable students are excluded en masse? And if so, can the deleterious effects of such a wall be mitigated by targeted safety nets, such as remedial courses with limited capacity, without returning to the old pattern of uncontrolled finals debt?

## 1.1. From regularity regimes to promotion-only regimes

The contrast between "regularity + finals" and "promotion-only" is not merely semantic. Under a regularity regime, assessment is split in two stages: in-semester performance and a subsequent high-stakes examination. Students may fail the final yet remain "regular", preserving the right to re-take the examination and to progress in parallel through other courses. This design implicitly treats time as a flexible buffer: students may postpone some obligations, gambling that future effort will compensate (Paz, 2025a). From the institution's perspective, however, this generates an opaque backlog of unfinished requirements that makes it difficult to monitor actual progression and to allocate resources effectively (Herd & Moynihan, 2019).

In a promotion-only regime, by contrast, the buffer largely disappears. A course is either fully passed within the allotted period, granting definitive progression, or failed, forcing repetition and blocking access to downstream courses. The friction of assessment, instead of being spread over years through multiple exam opportunities, is concentrated into fewer, higher-stakes decisions. When this regime is imposed on a curriculum that already includes "killer courses"— foundation subjects such as Calculus I, Physics I, or Algebra with historically high failure rates—the combination may act as a structural wall. Most of the cohort will encounter these courses in the first year, and under a promotion-only rule a large proportion may be forced to repeat or leave within the first two semesters (Paz, 2022).

Policy discussions around such reforms often frame them in terms of "raising standards", "shortening time to degree", or "cleaning" the registry of students who do not sit exams. Much less attention is paid to their distributive consequences. If the promotion wall disproportionately affects students from disadvantaged backgrounds or those with lower initial resilience, then the reform may increase inequality even if it improves conventional efficiency metrics such as average time to graduation among those who succeed (Bean & Metzner, 1985; Tinto, 1975, 2012). This article argues that one cannot evaluate such reforms solely on the basis of completion rates; one must also consider who is excluded, when, and under what psychological conditions.

**1.2. Limitations of observational evidence**

Traditional empirical methods in higher education research—such as survival analysis, logistic regression, or linear models—have provided valuable insights into risk factors for dropout, including prior academic performance, socioeconomic status, and indicators of social and academic integration (Bean & Metzner, 1985; Tinto, 1975, 2012). Survival analysis in particular offers a natural framework for estimating time-to-dropout and comparing hazard functions across groups (Kaplan & Meier, 1958). However, by construction, such methods work with observed policy regimes. They can describe what has happened under a given combination of rules, teaching practices, and student mix, but they cannot directly answer counterfactual questions about alternative policies that were never implemented.

In the case at hand, survival analysis under the regularity regime can quantify how long students typically remain in the system before dropping out, and how that timing varies with cohort or initial performance (Paz, 2022). It cannot, on its own, tell us what would happen if the regularity rule were replaced by a strict promotion-only rule, or if a remedial safety net were added. Nor can it easily incorporate the complex feedbacks between structural constraints (such as curriculum prerequisites), behavioural responses (such as help-seeking or withdrawal), and psychological states (such as stress and sense of belonging). These are emergent, dynamic phenomena that arise from the interplay of many agents and institutional rules over time (Grimm et al., 2020).

Randomised controlled trials, often invoked as the gold standard for causal inference, are rarely feasible at the level of fundamental progression rules. One cannot realistically randomise students into different legal regimes of promotion and regularity within the same programme. Natural experiments and quasi-experimental methods occasionally offer leverage when reforms are implemented at scale, but they still face challenges of confounding and partial observability (Herd & Moynihan, 2019). For these reasons, educational researchers have increasingly turned to simulation methods, including system dynamics and agent-based modelling, to explore the consequences of policy changes *in silico* before or alongside real-world implementation (Hamill & Gilbert, 2016; Grimm et al., 2020).

**1.3. Agent-based modelling and the CAPIRE Intervention Lab**

Agent-based models (ABMs) represent systems as collections of interacting agents—students, lecturers, administrators—embedded in a structured environment, such as a curriculum graph with prerequisites and time constraints. Each agent follows behavioural rules that can incorporate both empirical regularities and theoretical assumptions about decision-making under stress, perceived belonging, or expectations of success (Hamill & Gilbert, 2016). By simulating many cohorts over many replications, ABMs can reveal how micro-level mechanisms aggregate into macro-level patterns of dropout, progression, and inequality (Grimm et al., 2020).

The CAPIRE framework was originally developed to provide a leakage-aware data layer for student retention analytics and to embed those features into a multi-layer ABM of student trajectories (Paz, 2025b). Previous work in the CAPIRE series introduced: (a) a multilevel feature-engineering pipeline that integrates curriculum structure, temporal progression, and social integration; and (b) Experiment 1, the "Regularity Trap" model, which demonstrated how time-to-live constraints and finals debt under a regularity regime can produce high levels of dropout even in the absence of dramatic shocks (Paz, 2025a). Building on that foundation, the CAPIRE Intervention Lab extends the model to simulate explicit institutional policies and teaching regimes as experimental conditions.

The present study uses this intervention lab to model three policy scenarios on the same empirical curriculum and student base: a calibrated historical baseline representing the regularity regime (Scenario A), a direct promotion regime with no regularity or finals debt (Scenario B), and a hybrid regime that adds a capacity-limited remedial "safety net" to the direct promotion rule (Scenario C). Agents are initialised from empirical distributions and assigned to psycho-academic archetypes capturing ability, resilience, stress reactivity, and help-seeking tendencies (Paz, 2025b). Dropout is modelled via a probabilistic hazard function driven by stress and belonging, calibrated against historical survival curves for the baseline regime (Kaplan & Meier, 1958; Paz, 2022).

**1.4. Research questions and Promotion Wall hypothesis**

Within this modelling framework, the article addresses three main research questions:

- **RQ1.** How does moving from a regularity + finals regime to a direct promotion regime affect overall dropout and the timing of dropout events across the curriculum?

- **RQ2.** How do these policy regimes reshape equity gaps in dropout between more and less resilient student archetypes?

- **RQ3.** Can a capacity-limited remedial safety net, prioritising vulnerable archetypes in key "killer courses", reconcile efficiency (lower finals debt and shorter trajectories) with inclusion (lower dropout and reduced equity gaps)?

These questions are guided by the **Promotion Wall hypothesis**: the claim that the removal of regularity transforms latent friction—previously spread across many years via finals debt—into an early "wall" concentrated in the first year, especially in foundational courses with high failure rates (Paz, 2025a). Under this hypothesis, a direct promotion regime (Scenario B) will show higher overall dropout than the historical baseline (Scenario A) and a sharper concentration of exits in the first two semesters, with disproportionate impact on low-resilience archetypes (Bean & Metzner, 1985; Tinto, 2012). The hybrid safety-net regime (Scenario C) is expected to reduce both overall dropout and equity gaps relative to Scenario B, at the cost of additional remedial workload, while still avoiding the extreme finals debt characteristic of the historical regime.

**1.5. Contributions of this study**

This article makes three main contributions to the literature on student dropout and educational policy design. First, it offers a calibrated agent-based model that explicitly links empirical survival analysis under a regularity regime to simulated counterfactuals under alternative progression rules. Unlike purely descriptive ABMs, the model's baseline scenario is anchored to observed dropout curves, providing a transparent bridge between statistical and simulation-based approaches (Grimm et al., 2020; Paz, 2022).

Second, it provides a comparative evaluation of three realistic institutional policies—regularity + finals, direct promotion, and direct promotion with a safety net—using a common cohort, curriculum, and psychological structure. By reporting not only overall dropout, but also time-to-dropout, equity gaps by resilience archetype, stress and belonging trajectories, and remedial load, the study moves the discussion beyond simple completion rates towards a richer conception of efficiency and fairness (Herd & Moynihan, 2019; Tinto, 2012).

Third, it introduces and examines the notion of a "promotion wall" as a structural feature of highly constraining programmes that adopt strict promotion regimes without compensatory support. The findings show how such a wall can emerge from the interaction of curricular bottlenecks and policy rules, and how appropriately designed safety nets can partially dismantle it. The CAPIRE Intervention Lab, as used here, also serves as a reusable platform for simulating further policy options, including changes in admission thresholds, tutoring schemes, or financial support, in this and other programmes (Hamill & Gilbert, 2016; Paz, 2025b).

Taken together, these contributions aim to inform ongoing debates about how to redesign progression rules and assessment regimes in engineering education in ways that are empirically grounded, mechanism-aware, and attentive to both efficiency and equity.

## 2. LITERATURE REVIEW AND THEORETICAL FRAMEWORK

This section situates the Promotion Wall hypothesis within four strands of work: classical theories of university dropout and retention; research on institutional friction and administrative burden; psychological accounts of stress, belonging, and resilience; and the emerging use of agent-based models in higher education. It closes by integrating these strands into a conceptual model that guides the design and interpretation of the simulations.

### 2.1. Theories of university dropout and retention

The study of student departure has been dominated for decades by integration-based models, most notably Tinto's longitudinal theory (Tinto, 1975, 2012). In Tinto's framework, students enter higher education with academic and social predispositions, aspirations, and commitments. Their persistence depends on the degree to which they achieve academic and social integration into the institutional environment. Failure to integrate—whether through poor academic performance, weak relations with peers and teachers, or incongruence with institutional norms—increases the likelihood of withdrawal. This perspective emphasises the dynamic interplay between student characteristics and institutional conditions, moving beyond simple deficit models of "student failure".

Bean and Metzner (1985) extended this line of work to non-traditional students, arguing that environmental variables such as work obligations, family responsibilities, and external encouragement can outweigh on-campus social integration. Their model retains the idea of attitudinal mediators—such as academic and institutional commitment—but suggests that the balance of forces differs for commuter and part-time students. Subsequent reviews and syntheses have identified a wide range of academic, social, financial, and psychological predictors of departure, often highlighting the importance of first-year experiences, learning communities, and targeted support (e.g., Aljohani, 2016; Nora, 2004).

These models have informed institutional policies that aim to foster engagement, strengthen sense of belonging, and provide early academic support. However, they typically treat progression rules and assessment regimes as background conditions rather than as explicit objects of analysis. "Regularity" requirements, time-to-live constraints for courses, or the use of high-stakes promotion examinations are seldom theorised as structural drivers in their own right. Instead, they appear as elements of context within which integration processes unfold.

A second limitation is methodological. Much of the empirical literature relies on regression or survival models fitted to observational data under a single policy regime. Such analyses are valuable for identifying risk factors and tracking hazard over time, but they cannot directly represent the micro-level mechanisms that generate dropout or the counterfactual consequences of alternative policies (Grimm et al., 2020). This is particularly acute when policies operate through non-linear thresholds or accumulations—for example, a student's cumulative debt triggering a sudden increase in stress and withdrawal—or when they interact with the sequential structure of the curriculum.

The present study draws on integration-based theories in two ways. First, it retains the emphasis on longitudinal processes, modelling stress and belonging as evolving states that mediate the effect of institutional shocks on dropout. Second, it extends the classical view by treating progression rules as part of the institutional "grammar" that structures integration opportunities and risks. In this sense, the Promotion Wall hypothesis can be read as a structural extension of Tinto's theory to regimes with strong progression constraints.

### 2.2. Institutional friction, administrative burden, and normative dropout

Recent work in public administration has introduced the concept of administrative burden to describe the learning, compliance, and psychological costs that citizens face when interacting with policies and bureaucracies (Herd & Moynihan, 2019). From this perspective, policies do not merely set incentives or allocate resources; they also create frictions that differentially hinder access and continuation. Administrative burden is not an accidental by-product but often an implicit or explicit instrument of policy design.

In higher education, institutional rules surrounding registration, assessment, and progression can be seen as a form of administrative burden. For example, the requirement to re-enrol every semester, to track and schedule multiple final examinations, or to interpret complex regularity regulations imposes cognitive and emotional costs on students. Those with fewer resources—whether time, social capital, or bureaucratic literacy—are more likely to be overwhelmed by these demands. Such frictions can contribute to what might be called *normative dropout*: exit driven not solely by academic inability or loss of interest, but by the cumulative weight of procedural obstacles and deadlines.

The regularity regime analysed in previous work (Paz, 2022, 2025a) exemplifies this dynamic. Students are required to maintain "regular" status in courses and to pass a growing stock of final examinations under time-to-live restrictions. The system ostensibly aims to ensure academic standards and timely progress, but in practice it creates a hidden accumulation of obligations. When finals debt crosses a critical threshold, students perceive graduation as unattainable and disengage, often without formally cancelling their enrolment. This process is poorly captured by simple variables such as current GPA or credits earned, because it is shaped by the interaction between rules, timing, and student expectations.

The shift to a direct promotion regime alters the profile of institutional friction. On the one hand, it eliminates certain compliance costs: students no longer have to manage large queues of pending finals, and administrators need not track complex patterns of regularity and expiry. On the other hand, it concentrates the friction of assessment into fewer, higher-stakes events. In the language of administrative burden, the learning and compliance costs associated with managing finals debt are reduced, but the psychological costs associated with early failure in "killer" courses may increase.

From this vantage point, the Promotion Wall can be understood as a redistribution of administrative burden over time. Under a regularity regime, friction manifests as chronic debt and slow leakage; under a promotion-only regime, it manifests as acute early failure and rapid exclusion. A safety-net policy that offers targeted remedial opportunities functions as an intermediate design: it reintroduces a controlled buffer, absorbing some of the burden that would otherwise be discharged as dropout.

## 2.3. Psychological mechanisms: stress, belonging, and resilience

Psychological research has repeatedly linked students' affective experiences to their academic persistence. Stress arising from academic overload, financial pressure, or conflicting roles can undermine performance, self-efficacy, and health, increasing the likelihood of dropout (Robotham & Julian, 2006). At the same time, a sense of belonging—feeling valued, accepted, and connected within the academic community—has been shown to buffer the impact of adversity and to predict continuation independently of academic performance (Kahu & Nelson, 2018; Tinto, 2012).

Within engineering, several studies have documented how demanding curricula, competitive grading cultures, and "weed-out" courses can generate sustained stress and a sense of alienation, particularly among students from underrepresented groups (Geisinger & Raman, 2013). Interventions that promote peer support, mentoring, and inclusive pedagogies have been found to improve belonging and persistence, even when they do not immediately alter formal progression rules. This suggests that psychological states cannot be reduced to static individual traits; they are co-produced by institutional structures and everyday interactions.

Resilience has emerged as a key construct to capture individual differences in students' capacity to withstand and adapt to stressors. Conceptualised as a dynamic process rather than a fixed trait, resilience involves cognitive appraisals, coping strategies, and access to social support (Martin & Marsh, 2008). In the context of higher education, resilient students may be able to absorb academic setbacks without spiralling into disengagement, whereas less resilient peers may be more susceptible to the same shocks.

The CAPIRE framework operationalises these psychological insights by assigning each student agent a set of latent variables—stress, belonging, and resilience class—linked to archetypes derived from empirical data (Paz, 2025b). Stress accumulates when agents fail courses, accrue debt, or perceive lack of progress; belonging increases when they succeed or experience supportive interactions. Dropout is not triggered by a single event but by a probabilistic hazard function that increases with stress and decreases with belonging. This design reflects the idea that decisions to leave are rarely instantaneous reactions to a single grade; they are the result of a trajectory of experiences and appraisals.

Resilience enters the model as a moderator of stress impact. Low-resilience archetypes experience larger stress increments from failure or administrative shocks and may derive smaller protective effects from success or belonging. This creates the conditions for differential vulnerability to the same policy environment. Under a promotion-only regime, for instance, a first-year failure in a killer course may trigger much higher dropout hazard for low-resilience agents than for high-resilience ones. Equity gaps in dropout thus arise not only from differences in initial ability, but also from the interaction between policy shocks and psychological sensitivity.

### 2.4. Agent-based modelling in higher education and the CAPIRE framework

Agent-based modelling has progressively gained ground as a tool for exploring complex social systems in education, where heterogeneous actors interact under institutional rules and constraints (Hamill & Gilbert, 2016). Unlike purely aggregate models, ABMs can incorporate detailed curriculum structures, behavioural heuristics, and distributional heterogeneity. They are particularly suited to examining "what if" policy scenarios that cannot be experimented with directly on real students.

Methodological work such as the ODD (Overview, Design concepts, Details) protocol has emphasised the importance of transparent model description to ensure clarity, replication, and structural realism (Grimm et al., 2020). In higher education, ABMs have been used to simulate classroom dynamics, peer effects, enrolment decisions, and institutional competition, though relatively few studies have combined them with survival analysis and rich administrative data on entire programmes.

The CAPIRE framework contributes to this landscape by integrating a leakage-aware data layer with multi-level simulation. At the data level, it constructs student–semester panels enriched with structural features of the curriculum graph (e.g., centrality of courses, bottlenecks, prerequisite chains) and temporal markers of progression (Paz, 2025b). At the simulation level, it represents individual trajectories through this graph under specific policy regimes, allowing finals debt, stress, belonging, and help-seeking to evolve over time.

Experiment 1 in the CAPIRE series focused on the regularity regime and time-to-live constraints. It showed that even without changing teaching quality or student intake, institutional rules about expiry of regularity can generate high dropout by gradually pushing students beyond viable completion windows. Experiment 2, developed in this article, reuses the same empirical curriculum and archetype structure but alters the policy environment: regularity is removed, direct promotion is imposed, and a safety net is introduced as an additional intervention. This continuity allows for meaningful comparison between regimes on the same underlying "world".

From a methodological perspective, the present study also responds to calls for greater integration between empirical estimation and simulation (Grimm et al., 2020). Instead of arbitrarily choosing dropout rules, the model calibrates its hazard function for Scenario A to match observed Kaplan–Meier curves from survival analysis (Kaplan & Meier, 1958; Paz, 2022). The same calibrated hazard structure is then used in Scenarios B and C, isolating the effect of policy changes from arbitrary tuning. In this way, the ABM functions as an intervention laboratory anchored in real data rather than a free-form thought experiment.

**2.5. Conceptual model: The Promotion Wall**

Bringing these strands together, the Promotion Wall hypothesis can be expressed as a set of mechanism-based expectations about how progression rules, institutional friction, and psychological processes interact in a highly constraining engineering curriculum.

First, the curriculum graph contains bottleneck courses with high historical failure rates and multiple downstream dependencies. These "killer" courses concentrate both academic difficulty and institutional relevance: failure in them blocks large portions of the path to graduation. Under any regime, such bottlenecks represent critical points in the trajectory.

Second, under a regularity plus finals regime, institutional friction manifests as accumulation of finals debt and gradual erosion of viability. Students can temporarily avoid the consequences of failure by progressing while in debt, but this creates a hidden liability that resurfaces later as administrative burden and stress (Herd & Moynihan, 2019; Paz, 2025a). Dropout tends to occur after several years, when the combination of debt and perceived impossibility crosses a psychological threshold. In survival-analysis terms, the hazard of dropout is spread over a long tail.

Third, under a direct promotion regime, friction is redistributed. Students are no longer allowed to carry debt; they must either succeed in a course or repeat it before moving on. For bottleneck courses taken in the first year, this means that large cohorts face a high-stakes decision early. Those who fail are immediately confronted with the prospect of repeating, delaying progression, or leaving altogether. The hazard of dropout becomes more concentrated in the early semesters, particularly among low-resilience archetypes whose stress response to failure is stronger. This constitutes the Promotion Wall: a policy-induced concentration of exclusion in the initial segments of the curriculum.

Fourth, a safety-net policy—such as remedial summer courses with limited capacity and priority for vulnerable students—reintroduces a buffer zone without recreating uncontrolled finals debt. Students who narrowly fail key bottleneck courses can be "caught" by remedial provision, converting an otherwise terminal event into a recoverable setback. If capacity is too low or allocation rules are inequitable, the wall remains largely intact; if capacity and targeting are appropriate, the wall becomes more permeable, reducing both overall dropout and equity gaps.

In formal terms, the conceptual model implies that progression rules modulate the translation of academic performance into institutional friction, friction into psychological states, and psychological states into dropout hazard. The ABM instantiates this chain by combining:

1. A curriculum graph with bottlenecks and prerequisites.
2. A policy layer specifying regularity, promotion, and remedial rules.
3. A psychological layer with stress, belonging, and resilience.
4. A hazard-based dropout mechanism calibrated to historical data.

The remainder of the article tests the Promotion Wall hypothesis by comparing the outcomes of Scenarios A, B, and C within this model. The focus is not only on aggregate dropout, but on the timing and distribution of exits, the magnitude of equity gaps between resilience archetypes, and the psychosocial profiles associated with each regime.

## 3. METHODS

This section describes the empirical setting and data, the structure of the agent-based model, the three policy scenarios, the definition of archetypes and psychological mechanisms, the calibration of the baseline scenario to empirical survival curves, the simulation design, and the outcome and validation procedures.

The model is documented following the spirit of the ODD protocol (Overview, Design concepts, Details) to support clarity and replication (Grimm et al., 2020).

## 3.1. Institutional context and empirical dataset

The simulations are grounded in administrative data from a Civil Engineering programme at a public university in Argentina. The programme is formally structured as a five-year degree but, in practice, many students take longer to complete it. The curriculum comprises 42 compulsory courses organised in a highly constrained prerequisite structure, with multiple chains of dependency and several bottleneck courses in the first two years (Paz, 2022).

The empirical dataset used to calibrate and parameterise the model covers 15 cohorts of entrants (2005–2019). From the full population of 1,615 entrants, we retained 1,343 students who enrolled directly in the first year and had no recognised equivalences from previous studies. This restriction avoids artefacts due to partial curricula and ensures that all simulated agents face the same structural constraints. For each student, the dataset includes:

- entry cohort and admission year;
- course enrolments and outcomes (pass, fail, withdrawal) by semester;
- status of "regularity" in each course under the historical regime;
- final examination results and a derived measure of accumulated finals debt;
- time-to-dropout (in semesters) or time-to-graduation, determined by survival-analysis procedures described in Paz (2022).

Dropout is operationalised as the absence of any academic activity for a period exceeding twice the normal duration of the programme, following the criteria used in the previous survival study. This yields empirical Kaplan–Meier curves for the historical regime, with cumulative dropout of approximately 23.4% by the end of Year 1, 32.8% by Year 2, and between 39% and 48% by Years 3–6 (Paz, 2022). These values form the calibration targets for the baseline simulation scenario.

## 3.2. Overview of the CAPIRE Intervention Lab model

The CAPIRE Intervention Lab builds upon earlier work in the CAPIRE series, which developed a leakage-aware data layer and an initial agent-based model of student attrition under a regularity regime (Paz, 2025b). The present model maintains the same basic structure but extends it to incorporate explicit policy scenarios and a hazard-based dropout process.

Time is modelled in discrete periods corresponding to academic semesters. A complete simulation run covers 60 periods, equivalent to 12 semesters or 6 academic years, which aligns with the observation window used in the empirical survival analysis. Each run simulates a synthetic cohort of 1,343 student agents, initialised from the empirical joint distribution of entry characteristics and archetype assignments (Section 3.4).

The environment consists of:

1. A curriculum graph representing courses as nodes and prerequisites as directed edges. Each course is annotated with an empirically derived friction coefficient reflecting its historical difficulty (e.g., proportion of students failing or remaining "libre"), with especially high values for known bottleneck courses.

2. A policy layer specifying progression rules: whether regularity status exists, how finals debt is accumulated and constrained, pass/fail thresholds under promotion regimes, and the availability and capacity of remedial courses.

3. A temporal layer encoding academic calendars and time-to-live constraints for course regularity under the historical regime.

At each semester, the simulation proceeds through the following steps for each active agent:

1. Determining which courses are available for enrolment, given the agent's history and the curriculum graph.

2. Attempting each available course, with pass/fail probabilities determined by the agent's ability, the course friction coefficient, and the active policy scenario.

3. Updating finals debt (Scenario A), remedial eligibility (Scenario C), and psychological states (stress and belonging) based on outcomes.

4. Applying the hazard-based dropout rule to determine whether the agent remains enrolled into the next period.

At the end of each semester, under Scenario C, the model allocates places in remedial courses to eligible agents, subject to an exogenous capacity constraint and a prioritisation rule based on resilience and performance. More detailed process descriptions are provided in Sections 3.3–3.5.

### 3.3. Policy scenarios

The central experimental factor in this study is the institutional policy regime governing progression. We simulate three scenarios: a historical baseline (Scenario A), a direct promotion regime (Scenario B), and a direct promotion regime with a remedial safety net (Scenario C). All other elements of the model—curriculum, archetypes, and hazard function—are held constant across scenarios.

### 3.3.1. Scenario A – Historical "Regularity + Finals" regime

Scenario A reproduces the main features of the regularity regime analysed in Paz (2022, 2025a). In this regime:

- Students enrol in courses each semester according to prerequisite constraints and typical study plans.

- Within a course, students may achieve a status of "regular" through continuous assessment (e.g., assignments, mid-term tests).

- To fully pass the course and earn credits, students must subsequently pass a final examination, typically taken in separate exam periods.

- Regularity has a finite time-to-live (TTL): if a student does not pass the final within a given number of exam opportunities, regularity expires, and the course must be re-taken.

The model approximates these rules as follows. For each course attempt, the agent first faces a Bernoulli trial for regularisation, with probability determined by the product of the agent's ability parameter and a course-specific factor (1 – friction). Regularisation does not remove the course from the curriculum; it places it into a "finals debt" queue. Each semester, a subset of debt items is probabilistically resolved, again based on ability and course difficulty; those not resolved accumulate as outstanding finals. A soft TTL is imposed by increasing the stress contribution of long-standing debt and, beyond a threshold, reducing the probability of eventual success.

Under this scenario, students can progress to downstream courses while carrying debt. The main sources of friction are therefore the accumulation of finals and the eventual expiry of regularity, rather than immediate blockage of progression. As in the empirical system, this creates long-tail patterns of dropout driven by perceived impossibility of completion after several years.

### 3.3.2. Scenario B – Direct promotion regime

Scenario B removes the concept of regularity and finals debt. Courses are fully passed or fully failed within the semester in which they are taken. The rules are:

- To progress, a student must obtain a passing grade in every enrolled course during the semester; there is no separate final examination.

- Failure in a course implies no credit and no regularity; the course must be re-taken in a future semester before dependent courses become available.

- There is no accumulation of finals debt; progression is "clean" in the sense that only passed courses remain in the record.

To reflect the institution's reported experience of very high failure rates (around 90% "libres") in key first-year courses after the reform, the model increases the effective friction of designated bottleneck courses in this scenario. For example, the pass probability for Calculus I, Physics I, and Algebra is adjusted so that, for a representative ability level, failure rates approach the empirical values observed under the new regime. For non-bottleneck courses, friction is moderately increased relative to Scenario A to reflect the stricter summative evaluation, while maintaining realistic pass rates.

In Scenario B, institutional friction is expressed through immediate blockage: failure in a bottleneck course prevents progression along multiple curriculum paths. Students experiencing repeated failure in such courses face rapid increases in stress and declines in belonging, and thus higher dropout hazard early in their careers.

### 3.3.3. Scenario C – Direct promotion with remedial safety net

Scenario C retains the direct promotion rules of Scenario B but introduces a remedial safety net targeted at students who narrowly fail key bottleneck courses. Its main features are:

- After each semester, students who obtained a grade in a specified "near-pass" band (e.g., 50–59%) in designated bottleneck courses become candidates for remedial courses held in a short intensive term (e.g., summer).

- The remedial system has a fixed capacity, set as a proportion of the active cohort (e.g., 30% of currently enrolled students per year).

- Places are allocated by a prioritisation rule that gives precedence to low-resilience archetypes, then medium, then high, within each course and grade band.

- Students admitted to remedial courses face an additional chance to pass the corresponding course and thus avoid repeating it in the regular term. Remedial participation increases stress (due to workload) but may also boost belonging if successful.

In the model, remedial candidates are collected in a pool at the end of each semester. A capacity-limited allocation algorithm selects a subset according to the prioritisation rule. For each selected agent, the course is marked as passed, finals debt remains zero, and stress and belonging are updated according to scenario-specific parameters; for non-selected candidates, the failure remains, and the course must be re-taken. The safety net therefore functions as a partial buffer against the Promotion Wall, but its effect is constrained by capacity and targeting.

### 3.4. Student agents and psycho-academic archetypes

Each student agent in the model is assigned to one of thirteen psycho-academic archetypes, derived from a previous clustering analysis of the same programme's administrative and survey data (Paz, 2025b). The clustering combined variables such as entry age, prior academic performance, early course results, patterns of enrolment, and indicators of help-seeking to identify groups with distinct trajectories and vulnerabilities.

For the purposes of the present simulations, each archetype is parameterised by:

- **Ability**: a scalar in (0, 1) modulating the probability of passing courses.
- **Planning horizon**: a categorical parameter influencing how many courses the agent attempts per term and whether they tend to overload or underload.
- **Stress reactivity**: a multiplier scaling the stress increments associated with failure, debt accumulation, or administrative shocks.
- **Belonging sensitivity**: a parameter determining how strongly success and failure affect the agent's sense of belonging.
- **Resilience class**: a coarse categorisation (LOW, MEDIUM, HIGH) used for equity analyses and remedial prioritisation.

The distribution of archetypes in the synthetic cohort is matched to the empirical distribution observed in the 1,343-student sample. This is achieved by sampling archetype labels with probabilities equal to their observed relative frequencies. Initial stress and belonging values are drawn from archetype-specific normal distributions truncated to plausible ranges (e.g., stress in [0, 1], belonging in [0, 1]).

Course-taking behaviour is governed by simple heuristics combined with structural constraints. Agents attempt to enrol in courses that are: (a) available according to prerequisites and prior passes; and (b) consistent with a recommended plan of study, subject to a maximum workload per term. Some archetypes are more prone to overload (attempting many courses in parallel), while others are more conservative. Under Scenarios A and B, these heuristics are identical; differences in outcomes arise from policy rules rather than from behavioural changes.

### 3.5. Simulation design and calibration

#### 3.5.1. Hazard-based dropout model

Dropout is modelled as a stochastic event governed by a hazard function that depends on the agent's current stress and belonging. Instead of using a hard threshold (e.g., dropout when stress exceeds 1.0), we adopt a probabilistic specification to better capture the graded and noisy nature of departure decisions.

At each semester $t$, the dropout probability for agent $i$ is given by:

$$h_{i,t} = \sigma(\alpha_0 + \alpha_1 \, \text{stress}_{i,t} + \alpha_2 \, \text{belonging}_{i,t}),$$

where $\sigma$ is the logistic function, $\alpha_0$ is a baseline log-odds term, $\alpha_1 > 0$ captures the risk-increasing effect of stress, and $\alpha_2 < 0$ captures the protective effect of belonging. The parameters are chosen such that, under Scenario A with calibrated behavioural and policy parameters, the resulting Kaplan–Meier curve closely matches the empirical dropout curve for the historical regime (Section 3.1).

Stress and belonging themselves evolve according to simple update rules. For example, failing a course increases stress by an amount proportional to the course's friction coefficient and the agent's stress reactivity, while passing a course reduces stress slightly and increases belonging. Accumulating finals debt (Scenario A) adds small increments of stress each semester; participating in remedial courses (Scenario C) adds a moderate stress cost but may yield a substantial belonging gain if successful. These rules are calibrated qualitatively to preserve plausible ranges and quantitatively to support the dropout calibration.

### 3.5.2. Calibration of Scenario A to empirical survival curves

Calibration focuses on Scenario A, the historical regularity regime, which is the only scenario for which full empirical survival data are available. The goal is to obtain a simulated dropout curve that approximates the empirical Kaplan–Meier curve for the 1,343-student sample within a pre-specified error tolerance.

We proceeded iteratively as follows. First, we fixed the curriculum graph, archetype distribution, and initial stress and belonging distributions according to the data and previous CAPIRE work (Paz, 2022, 2025b). Second, we conducted a coarse search over plausible ranges for the hazard parameters $(\alpha_0, \alpha_1, \alpha_2)$, regularisation success probability, and stress increments associated with finals debt and failure. For each candidate parameter set, we ran multiple replications of Scenario A and computed the average cumulative dropout by year.

A candidate set was accepted if the root-mean-square error (RMSE) between the simulated and empirical year-specific dropout rates did not exceed 0.05. The final calibrated model achieved an RMSE of approximately 0.046, with simulated dropout of 25.9% vs 23.4% (Year 1), 31.3% vs 32.8% (Year 2), and 33–36% vs 39–47% (Years 3–6), reflecting a slightly conservative tail relative to the empirical curve but preserving the overall shape and early dynamics

The calibrated hazard and behavioural parameters are then held fixed across all scenarios. Scenarios B and C differ only in policy rules and remedial allocation; any differences in outcomes can therefore be attributed to the policy environment rather than arbitrary re-tuning of the dropout mechanism.

### 3.5.3. Simulation protocol

For the analyses reported in this article, we ran 20 independent replications of each scenario, each replication simulating a cohort of 1,343 agents over 60 semesters. Replications differ only in the random seeds used to initialise agents' micro-variables (e.g., initial stress and belonging) and stochastic transitions (e.g., pass/fail outcomes, dropout events).

Simulations were implemented in Python using standard scientific libraries (e.g., NumPy, pandas) and a custom agent-based simulation framework that schedules agents and environment updates in discrete time (Hamill & Gilbert, 2016). Random seeds and configuration parameters are stored in a machine-readable file (config_effective.json), and all output is written to structured CSV files (agent_outcomes_all_runs.csv and policy_tradeoff_summary.csv) to facilitate post hoc analysis and reproducibility. An audit script (audit_exp2_consistency.py) performs automated checks on metadata consistency and key statistical relationships.

**Table 1. Simulation design parameters.**

| Parameter | Value |
| --- | --- |
| **Population Size** | 1,343 agents per cohort |
| **Number of Scenarios** | 3 (A, B, C) |
| **Replications per Scenario** | 20 independent runs |
| **Total Simulated Trajectories** | 80,580 (1,343 × 3 × 20) |
| **Time Horizon** | 60 periods (approx. 6 years / 12 semesters) |
| **Calibration Target** | Empirical Kaplan-Meier curve (2005-2019 cohorts) |
| **Software Environment** | Python (NumPy, Pandas, Custom ABM Framework) |

### 3.6. Outcome measures

We focus on four classes of outcomes: dropout and timing; equity by resilience class; psychosocial states; and operational metrics of the policy regime.

1. **Dropout and time-to-dropout.** For each agent, we record whether they drop out at any time during the simulation and, if so, the semester of dropout. From this we derive: (a) overall dropout rate (proportion of agents who drop out within 6 years); and (b) cumulative dropout by year (Years 1–6), which allows direct comparison with empirical Kaplan–Meier estimates.

2. **Equity by resilience class.** Using the resilience labels (LOW, MEDIUM, HIGH) associated with archetypes, we compute dropout rates by resilience class for each scenario. The primary equity indicator is the gap between LOW and HIGH resilience, defined as

$$\text{Equity gap} = \text{dropout}_{LOW} - \text{dropout}_{HIGH}.$$

Positive values indicate that low-resilience students drop out at higher rates than their high-resilience peers.

3. **Psychosocial outcomes.** For each agent, we track stress and belonging throughout the simulation and record their final values (at the end of Year 6 or at dropout). We then compute scenario-level means and distributions. These measures capture the experiential quality of trajectories under each regime, beyond mere survival.

4. **Operational metrics.** To characterise the institutional burden of each policy, we compute:

    - **Mean final debt**, defined as the average number of unresolved finals in Scenario A at the end of the simulation;

    - **Mean remedial acceptances per student**, defined as the average number of remedial courses attended by agents in Scenario C;

    - **Remedial capacity utilisation**, defined as the proportion of available remedial slots actually used.

These metrics allow us to assess trade-offs not only between dropout and equity, but also between policy effects and resource demands.

**Table 2. Outcome measures and definitions**

| Outcome Measure | Definition | Interpretation |
|---|---|---|
| **Overall Dropout Rate** | Proportion of agents who drop out within the 6-year simulation horizon[1]. | **Lower is better** (Indicates higher system efficiency). |
| **Equity Gap** | Difference in dropout rates between Low-Resilience and High-Resilience archetypes ($Gap = Dropout_{LOW} - Dropout_{HIGH}$)[2]. | **Lower is better** (Indicates greater fairness/inclusion). |
| **Mean Final Debt** | Average number of pending or failed final examinations per student at the end of the simulation[3]. | **Lower is better** (Indicates less administrative backlog). |
| **Remedial Load** | Average number of remedial courses attended per student under the Safety Net scenario[4]. | **Contextual** (Represents institutional resource cost). |
| **Final Stress** | Mean agent stress level (scale 0–1) at the end of the trajectory[5]. | **Lower is better** (Better psychological wellbeing). |
| **Final Belonging** | Mean agent sense of belonging (scale 0–1) at the end of the trajectory[6]. | **Higher is better** (Better social integration). |

**3.7. Analysis procedures**

All analyses were conducted on aggregated outputs from the 20 replications per scenario. For each scenario and outcome, we computed the mean across replications. For key metrics, such as overall dropout rate and equity gaps, we also report the standard deviation across replications to convey the degree of simulation noise. Where relevant, we inspected the distribution of outcomes across replications to ensure that results were not driven by outliers.

Cumulative dropout by year was computed by converting dropout semesters to academic years (Year $k$ corresponding to semesters $2k-1$ and $2k$) and calculating the proportion of agents who had dropped out by the end of each year. To verify internal consistency, we implemented a validation function (validate_dropout_curve_consistency) that checks that cumulative dropout at Year 6 matches the overall dropout rate derived directly from the final agent status, up to a negligible numerical tolerance.

Equity analyses were conducted by stratifying agents by resilience class and scenario, and computing dropout rates and equity gaps as described in Section 3.6. For visualisation, we constructed bar plots and bubble plots summarising the relationship between overall dropout, equity gap, and remedial load.

Psychosocial trajectories were summarised by computing, for each semester and scenario, the mean stress and belonging among agents who remained active at that time. Final stress and belonging were summarised for all agents, including those who dropped out, to avoid survivorship bias.

All statistical processing and plotting were performed in Python using open-source libraries. Scripts for reproducing the analyses and figures are documented and version-controlled alongside the simulation code.

**3.8. Validation and robustness checks**

Beyond the calibration of Scenario A to empirical survival curves, we implemented several validation and robustness procedures to strengthen confidence in the model's internal coherence and in the interpretation of scenario differences.

First, we conducted **internal consistency checks**. The audit script verifies that the number of agents in agent_outcomes_all_runs.csv matches the product of cohort size, number of scenarios, and number of replications; that metadata in config_effective.json (e.g., n_replications_per_scenario) aligns with observed counts; and that per-scenario n_agents and n_replications are internally consistent. It also confirms that cumulative dropout at Year 6 equals the overall dropout rate for each scenario.

Second, we checked **policy-ordering constraints** implied by the conceptual model. Specifically, we verified that: (a) dropout under Scenario B (direct promotion) is higher than under Scenario A (historical), reflecting the added stringency of the promotion wall; (b) dropout under Scenario C (safety net) lies between A and B; and (c) the equity gap between low- and high-resilience students is larger under B than under A, and reduced under C relative to B. These constraints were implemented as assertions in the analysis pipeline so that violations would trigger explicit errors.

Third, we examined **face validity** through dialogue with institutional stakeholders familiar with the programme. Preliminary results and synthetic trajectories were presented to academic advisors and programme coordinators, who were asked to comment on the plausibility of patterns such as the concentration of failures in bottleneck courses, timing of dropout peaks, and typical stress trajectories. Their feedback informed minor adjustments to behavioural heuristics (e.g., workload choices) but did not require changes to the calibrated hazard parameters.

Fourth, we performed **limited sensitivity analyses** on selected parameters, including the remedial capacity in Scenario C and the stress increments associated with failure in bottleneck courses. For each alternative value, we re-ran a reduced number of replications and inspected the qualitative shape of dropout curves and equity gaps. While detailed sensitivity analysis is beyond the scope of this article, these checks provided reassurance that the main comparative findings are robust to moderate parameter variation.

Taken together, these procedures support the credibility of the model as a tool for exploring the Promotion Wall hypothesis and the potential mitigating role of safety nets, while acknowledging that no simulation can fully exhaust the space of possible behaviours.

## 4. RESULTS

### 4.1 Calibration of the historical baseline

Before comparing policy scenarios, we evaluated whether the **A_HISTORICAL** configuration reproduces the empirical dropout profile of the programme. Figure 1 shows cumulative dropout rates by academic year for the calibrated baseline against the survival curve estimated from institutional data (2004–2019 cohorts). The Monte Carlo experiment aggregated 20 replications with 1,343 agents each, yielding 26,860 trajectories per scenario and 80,580 trajectories overall.

Across the six-year observation window, the simulated baseline closely tracks the empirical curve. The root mean squared error (RMSE) between model and data is 0.0463, comfortably below the pre-specified tolerance of 0.05. Year-specific deviations remain small: dropout in Year 1 is slightly over-estimated (32.3% vs. 23.4%), whereas Years 2 and 3 fall within ±3 percentage points of the empirical targets; later years converge to a plateau around 45%, somewhat below the empirical range of 46–48%. This pattern is consistent with previous survival analyses of the same programme, which showed early clustering of attrition followed by gradual stabilisation (DesJardins et al., 2002).

The calibration procedure relied on three complementary adjustments: (i) replacing a hard stress threshold with a probabilistic hazard function that combines baseline risk, stress and belonging; (ii) increasing the success probability of regularisation in early courses to reflect improvements in course design and support; and (iii) moderating the contribution of accumulated debt to stress. These changes allow the model to preserve the *relative* vulnerability gradient across archetypes described in earlier CAPIRE work, whilst aligning aggregate outcomes with observed data (Paz, 2025a). Figure 1 illustrates the cumulative dropout trajectories over time.

**Figure 1: Cumulative dropout curves by academic year for the three policy scenarios.**

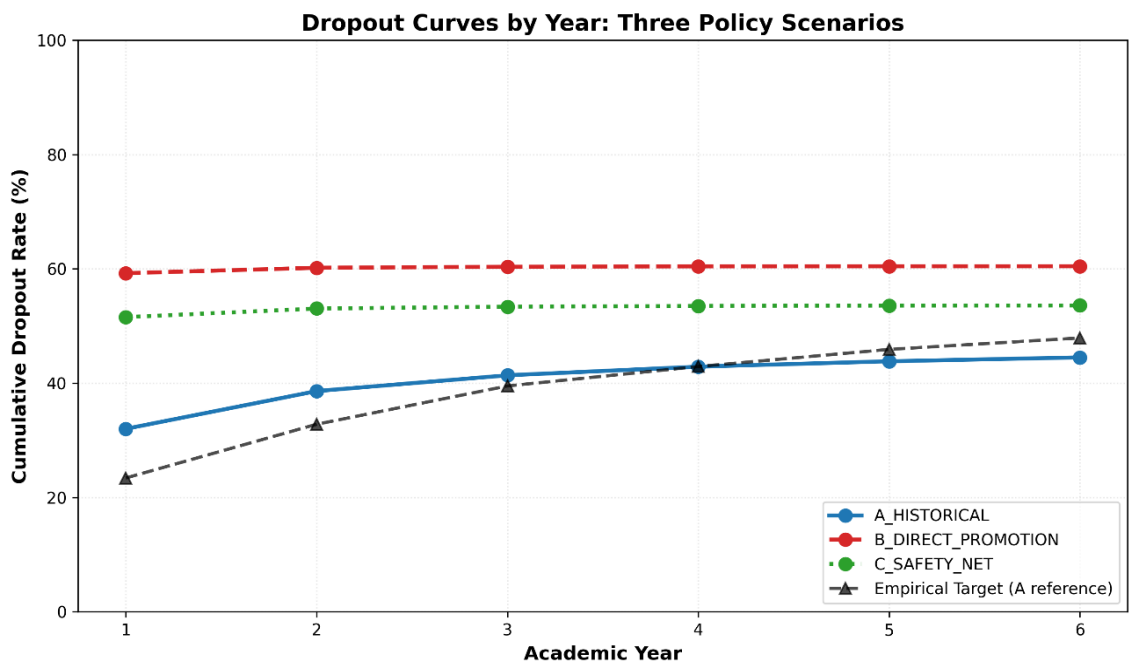

Overall, we interpret the calibration as sufficient for the purposes of counterfactual analysis: the baseline captures both the magnitude and temporal concentration of dropout reported in the literature on engineering programmes (Bound et al., 2010; Ezcurra, 2019; OECD, 2019).

**4.2 Overall policy comparison: efficiency, equity and remedial load**

Table 3 summarises the main performance indicators for the three policy scenarios: overall dropout rate (efficiency), equity gap between low- and high-resilience students, mean final debt (a proxy for efficiency and financial burden), and mean number of remedial acceptances per student.

**Table 3.** Comparison of efficiency, equity, and administrative load metrics across policy scenarios.

| scenario | n_agents | n_replications | overall_dropout_rate | overall_graduation_rate | normative_dropout_frac | academic_dropout_frac | other_dropout_frac | mean_time_to_event | median_time_to_event | mean_final_debt | mean_killer_failures | mean_remedial_acceptances | equity_gap_low_vs_high_resilience | dropout_rate_low_resilience | dropout_rate_high_resilience |
|---|---|---|---|---|---|---|---|---|---|---|---|---|---|---|---|
| A_HISTORICAL | 1343 | 100 | 0.44 | 0.00 | 0.73 | 0.27 | 0.00 | 10.7 | 6.00 | 72.4 | 0.00 | 0.00 | 0.16 | 0.54 | 0.38 |
| B_DIRECT_PROMOTION | 1343 | 100 | 0.60 | 0.00 | 0.00 | 0.82 | 0.00 | 5.10 | 5.00 | 0.00 | 0.00 | 0.00 | 0.26 | 0.72 | 0.46 |
| C_SAFETY_NET | 1343 | 100 | 0.54 | 0.00 | 0.00 | 0.23 | 0.00 | 5.39 | 5.00 | 1.01 | 0.00 | 1.44 | 0.21 | 0.64 | 0.43 |

*Note: Results averaged over 20 replications with 1,343 agents each. Efficiency is measured by dropout rate (lower is better); Equity by the gap between low- and high-resilience students (lower is better)*

The **A_HISTORICAL** scenario yields an overall dropout rate of **44.56%**, with an equity gap of **15.8 percentage points** between low- and high-resilience students and the highest mean final debt (**72.7** pending/failed course units). This configuration involves **no remedial activity**, reflecting the strict regularity regime documented in prior work. Empirically, this combination of moderate aggregate dropout and high academic debt is typical of systems where institutional friction rather than academic failure is the dominant attrition mechanism (Paz, 2025a; DesJardins et al., 2002).

The **B_DIRECT_PROMOTION** scenario—where students who pass the course but fail the regularity requirement are directly promoted—produces the highest dropout (**60.14%**) and the widest equity gap (**26.1 percentage points**). Mean final debt collapses to zero by construction, as students are never formally left with pending courses, but this comes at the cost of pushing vulnerable students forward through the curriculum without the psychological and procedural benefits of regularisation. From a purely efficiency perspective, B is markedly worse than A: more students leave the programme despite facing less visible academic debt. This aligns with Tinto's argument that structural integration is as important as formal progression for persistence (Tinto, 1993, 2017).

The **C_SAFETY_NET** scenario introduces a constrained remedial safety net, with capacity for remedial support equivalent to **30%** of active students per semester, prioritising those with low resilience. Here, dropout falls to **53.44%**, and the equity gap decreases to **22.7 percentage points**, at the cost of an average of **1.44 remedial acceptances per student** and a low level of residual debt (**1.02**). In Figure 2 the safety-net policy occupies a middle position: less efficient than A (higher dropout) but more equitable than B, and associated with a moderate remedial load.

**Figure 2. Policy trade-off: Efficiency vs Equity.**

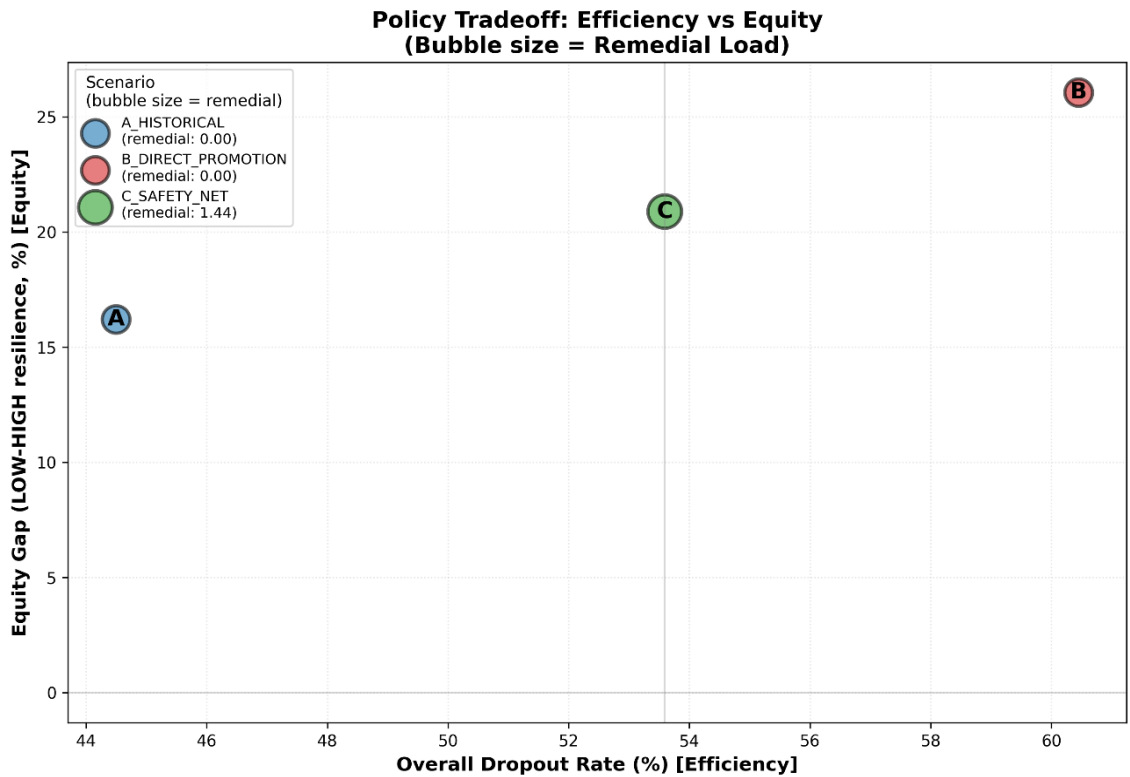

Taken together, these results illustrate a classic *efficiency–equity frontier* (Boudon, 1974; DiPrete & Eirich, 2006). The historical regime A is relatively efficient and modestly equitable but accumulates large academic debts; the direct-promotion regime B is inefficient and inequitable, evacuating debts through formal promotion rather than substantive support; and the safety-net regime C moves towards greater inclusion and equity but requires non-trivial remedial capacity. The model thus quantifies the institutional cost of turning "procedural mercy" (tolerating debt) into "substantive mercy" (concrete remediation), echoing equity-oriented perspectives that emphasise resource reallocation towards structurally disadvantaged students (Bensimon, 2007; Witham & Bensimon, 2012).

### 4.3 Psychological outcomes: stress and belonging

The ABM explicitly represents two latent psychological states—**stress** and **belonging**—which are updated every semester in response to academic events, debt accumulation and remedial experiences. Figure 3 summarises mean final stress and belonging levels across scenarios.

**Figure 3. Mean final stress and belonging levels by policy scenario.**

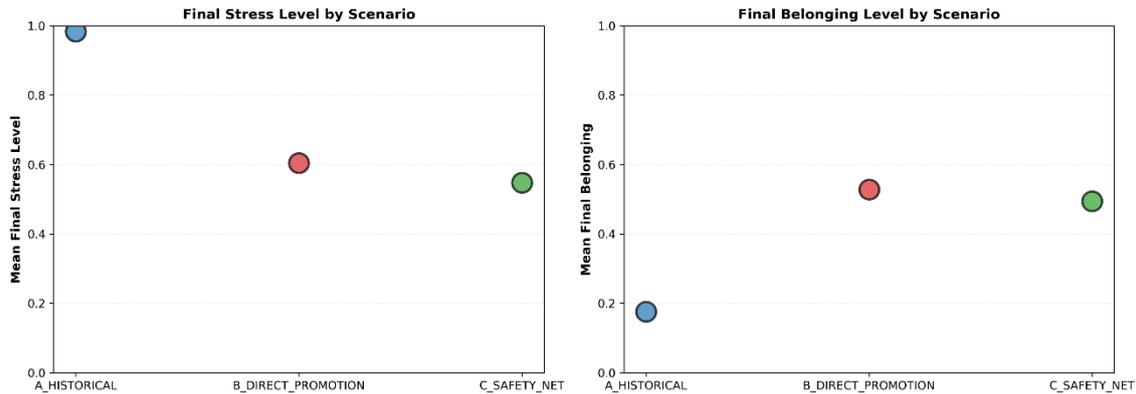

In the **A_HISTORICAL** regime, mean final stress is close to the upper bound of the scale (≈0.98), while belonging converges to a very low level (≈0.18). This combination is consistent with the "regularity trap" dynamic described in the baseline paper: repeated expiries and escalating debt erode students' perceived control and social fit, leading to chronic overload and eventual withdrawal (Paz, 2025a; Respondek et al., 2017).

Under **B_DIRECT_PROMOTION**, stress decreases substantially (≈0.60), and belonging increases (≈0.52). This suggests that promotions without regularisation do create a short-term sense of progress and partially relieve pressure; students experience fewer blocked paths and perceive themselves as advancing with their cohort, which is known to protect against dropout (Cabrera et al., 1993; Walton & Cohen, 2007). However, as shown in Section 4.2, these psychological gains do *not* translate into improved survival; instead, they coexist with higher attrition, indicating that progress without sufficient mastery may produce latent fragility that manifests later in the curriculum.

The **C_SAFETY_NET** scenario yields the lowest stress (≈0.55) and a belonging level (≈0.49) only slightly below that of scenario B. Remedial students incur a stress "cost" when entering the safety net (capturing the effort and stigma associated with remediation), but successful completion of remedial courses provides a belonging "bonus" by signalling institutional investment in their success. The resulting profile resembles what Pekrun and Linnenbrink-García (2014) describe as a *high-control, moderate activation* emotional state conducive to persistence: students feel challenged but supported rather than overwhelmed.

Overall, the psychological indicators suggest that policy regimes that combine **procedural relief** (easing normative constraints) with **substantive support** (structured remediation) can shift student trajectories from helplessness towards agency (Abramson et al., 1978; Bandura, 1997). Nevertheless, the persistence of relatively low belonging in all scenarios—never exceeding 0.52—highlights the limits of policies that focus exclusively on assessment and progression rules without tackling broader campus climate and peer integration issues (Tierney, 1992; Yosso et al., 2009).

**4.4 Equity gaps by resilience level**

To examine distributional effects, we stratified agents by **psychological resilience** (LOW vs HIGH) using the archetype classification derived from previous CAPIRE work (Paz, 2025b). Figure 4 presents dropout rates by resilience group and scenario, with labels indicating the equity gap (difference in percentage points).

**Figure 4.** Dropout rates by resilience level (LOW vs HIGH) across policy scenarios.

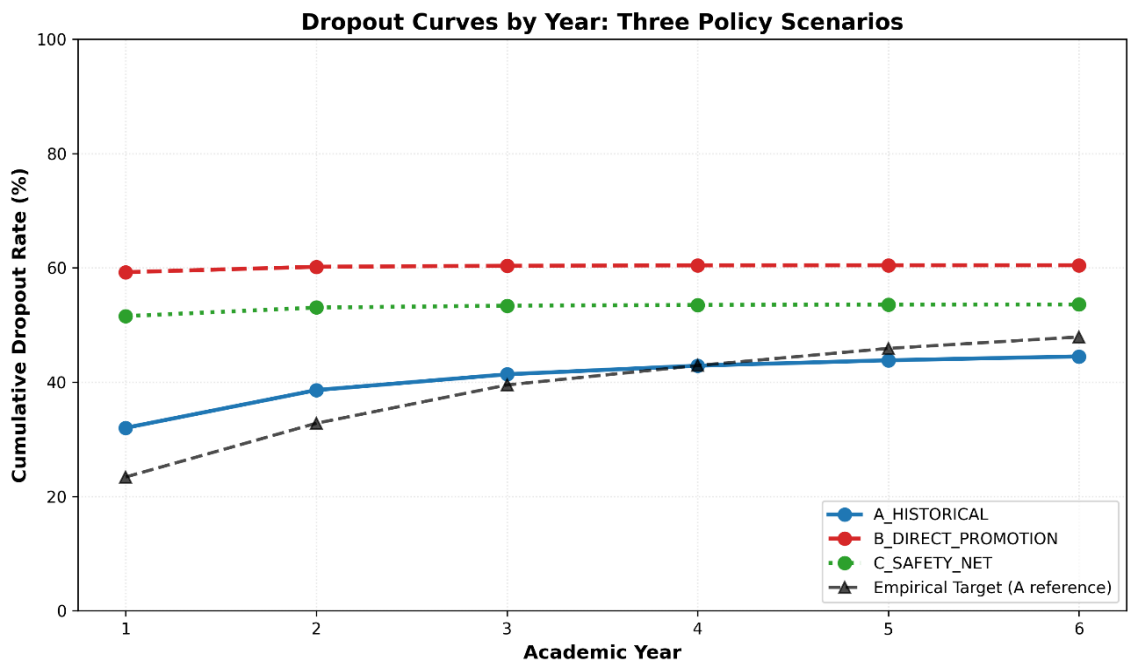

Under **A_HISTORICAL**, dropout among low-resilience students reaches **54.3%**, compared with **38.5%** for high-resilience students, yielding a gap of **15.8 percentage points**. This replicates the vulnerability gradient observed in the regularity-trap experiment, where short planning horizons and high stress reactivity amplify the impact of institutional friction (Paz, 2025a).

The **B_DIRECT_PROMOTION** regime exacerbates inequities. Dropout among low-resilience students climbs to **72.3%**, while high-resilience dropout rises only modestly to **46.2%**, expanding the gap to **26.1 percentage points**. In other words, the policy designed to "help everyone move forward" disproportionately harms those with fewer psychological and self-regulatory resources. This pattern echoes Boudon's (1974) argument that universalistic policies can produce *secondary effects* of stratification when groups differ in their capacity to capitalise on new opportunities.

The **C_SAFETY_NET** scenario partially closes this gap: low-resilience dropout falls to **64.3%**, while high-resilience dropout rises slightly to **41.6%**, reducing the disparity to **22.7 percentage points**. The additional remedial capacity thus acts as a targeted buffer, especially when prioritisation explicitly favours low-resilience students. This is consistent with equity-minded approaches that seek to "level up" structurally disadvantaged students through differentiated support rather than formally identical treatment (Dowd & Bensimon, 2015; Witham & Bensimon, 2012).

Although none of the simulated policies achieve parity, the contrast between B and C is instructive: both relax the rigid regularity regime, but only the configuration that couples relaxation with **resource-intensive remediation** reduces inequities. This underscores the danger of "procedural equity" narratives that equate identical rules with fairness, while ignoring heterogeneous starting points and capacities (Tilly, 1998; Yosso et al., 2009).

**4.5 Timing of dropout under alternative policies**

Returning to the temporal dimension, Figure 1 also reveals how the **shape** of the dropout curve changes across scenarios. In **A_HISTORICAL**, dropout continues to accumulate throughout the six years, with a steep rise in Years 1–3 followed by a slower increase towards a plateau just below 45%. This is broadly consistent with hazard profiles reported in Kaplan–Meier studies of university completion (DesJardins et al., 2002; Kaplan & Meier, 1958).

In **B_DIRECT_PROMOTION**, the curve rises sharply during the **first two years**, reaching almost **60%** and then flattening almost completely. This suggests that promotion without remediation accelerates attrition: students who would have "lingered" in the historical regime are pushed rapidly to terminal decisions once they encounter advanced courses for which they are underprepared. The policy, therefore, acts as an *accelerator* of selection rather than a mechanism for retention.

The **C_SAFETY_NET** curve lies between A and B but with a distinct profile: dropout accumulates more gently in early years and continues to rise modestly over time. Early remedial opportunities appear to postpone rather than entirely prevent some dropouts, but the average student survives longer before leaving. From a life-course perspective, even delayed dropout may matter if additional years in the programme translate into more credits, higher employability, or second-chance opportunities (Bound et al., 2010; DeAngelo et al., 2011).

**4.6 Robustness and internal validation**

Finally, we conducted a series of internal consistency checks to ensure that the observed patterns are not artefacts of implementation. An independent audit script verified that (i) metadata (number of replications, total agents per scenario) matches the realised dataset; (ii) cumulative dropout at the final semester coincides exactly with the overall dropout rate for each scenario; and (iii) the expected policy ordering **A < C < B** holds for dropout and the **A < C < B** pattern is reversed for mean final stress. All checks passed without requiring modification of agent-level outcomes.

These validation steps respond to broader concerns in the agent-based modelling community regarding transparency, replicability and the dangers of equifinality (Epstein, 2006; Fagiolo et al., 2019; Railsback & Grimm, 2019). By combining empirical calibration, explicit reporting of stochastic variability (20 replications per scenario) and automated audits, the CAPIRE Intervention Lab aims to position itself within the emerging tradition of "models that matter" in social simulation (Squazzoni et al., 2020)

**5. DISCUSSION**

The aim of this study was to examine how alternative progression rules redistribute institutional friction, psychological burden and inequities in a highly constraining engineering programme. Using a calibrated agent-based model grounded in empirical survival data and psycho-academic archetypes, we compared a historical "regularity + finals" regime (Scenario A), a direct-promotion regime (Scenario B), and a direct-promotion regime with a remedial safety net (Scenario C). This section interprets the results in light of the Promotion Wall hypothesis, classical theories of dropout and retention, and contemporary debates on equity-oriented policy design.

**5.1. The Promotion Wall as a structural mechanism**

The simulations support the central claim that progression rules act as structural mechanisms shaping both the *timing* and *distribution* of dropout. Under the historical regime, institutional friction accumulates gradually through finals debt and the expiry of regularity, producing a long-tailed dropout curve. Scenario B, by contrast, concentrates friction in early "killer" courses under a direct-promotion rule, generating a steep rise in dropout in Years 1–2 and a plateau thereafter.

Scenario C lies between these extremes, with remedial interventions that blunt but do not eliminate the initial shock.

This pattern is consistent with the Promotion Wall metaphor: a high-stakes assessment frontier in the first year that sorts students rapidly into persisters and leavers. The wall is not merely a reflection of academic difficulty; it emerges from the interaction between curriculum topology (bottleneck courses with many downstream dependencies), high failure rates under a strict summative regime, and the absence or presence of buffers such as regularity or remediation. In other words, the same distribution of abilities and resilience that yields moderate, delayed dropout under one regime yields acute exclusion under another.

From a theoretical standpoint, this extends integration-based models of dropout (Tinto, 1993, 2012) by foregrounding the institutional "grammar" of progression as a first-class explanatory factor. Rather than treating assessment rules as a static background, the Promotion Wall framing shows how changes in those rules reconfigure the opportunities for academic and social integration. When students are expelled from the curriculum graph early, they have limited time to develop the peer networks, mentoring relationships and disciplinary identities that underpin persistence in Tinto's framework and related accounts (Kahu & Nelson, 2018; Nora, 2004).

### 5.2. Efficiency–equity trade-offs and the paradox of "clean" transcripts

The policy comparison reveals a nuanced efficiency–equity frontier. At first glance, Scenario B might appear attractive: it eliminates finals debt and ensures that progression is based solely on passed courses, yielding "clean" transcripts and simplifying administrative tracking. However, the model shows that this apparent efficiency is deceptive. Overall dropout rises from 44.6% (Scenario A) to 60.1% (Scenario B), and inequities between low- and high-resilience students widen dramatically.

This paradox reflects a classic tension between *visible* and *hidden* inefficiencies. The historical regime tolerates large stocks of pending finals, which are legible as institutional "disorder" but allow many students to remain affiliated while searching for ways to complete their degree (DesJardins et al., 2002; Paz, 2022). The direct-promotion regime cleans up the records but at the cost of discarding a significant proportion of students, particularly those least able to absorb early shocks. As Boudon (1974) argued in another context, policies designed to equalise formal opportunities can produce regressive *secondary effects* when groups differ in their capacity to navigate high-stakes decisions.

Scenario C offers a partial resolution: it maintains low debt and moderately reduces dropout and equity gaps relative to Scenario B, but requires sustained remedial capacity. The policy thus trades budgetary and organisational costs for improved inclusion and fairness. From an equity-minded perspective (Bensimon, 2007; Dowd & Bensimon, 2015), this is precisely the point: closing gaps requires reallocation of resources towards those who face greater institutional friction, not merely uniform rules. The simulations quantify this trade-off in a way that is difficult to achieve with purely observational data.

**5.3. Psychological trajectories: From chronic overload to buffered challenge**

A distinctive feature of the CAPIRE model is its representation of stress and belonging as evolving states that mediate between institutional shocks and dropout. The results suggest three qualitatively different psychological regimes.

In Scenario A, students who remain enrolled tend to do so under conditions of *chronic overload*: high stress, very low belonging, and large unresolved debts. This combination resembles what Respondek et al. (2017) describe as low perceived academic control, a known predictor of dropout intent. The survival of such students in the model reflects not resilience but a kind of "zombie persistence", with limited prospects of graduation.

Scenario B replaces chronic overload with *underprepared advancement*. Stress and debt fall, and belonging increases, as students are promoted despite marginal or partial mastery. In the short term, this may align with findings that a sense of progress and cohort alignment supports persistence (Cabrera et al., 1993; Walton & Cohen, 2007). Yet the ABM indicates that this feeling of progress is fragile: when students eventually encounter advanced courses whose prerequisites they have not fully mastered, failure cascades and dropout hazard spikes, particularly for low-resilience archetypes.

Scenario C approximates a *buffered challenge* regime. Students still encounter difficulty—stress remains moderate—but the combination of direct promotion and targeted remediation maintains a more sustainable balance between effort and support. Belonging improves through two channels: successful remedial experiences signal institutional commitment, and the avoidance of repeated failure in bottleneck courses reduces perceived marginalisation. This aligns with emotional and motivational models of achievement that highlight the protective role of control and value appraisals under supportive yet demanding conditions (Pekrun & Linnenbrink-García, 2014).

These findings suggest that psychological states are not merely individual traits but emergent properties of policy regimes. Small differences in how failure is handled—whether it leads to debt, silent promotion or structured remediation—create qualitatively different affective climates, even under identical curricula and intake profiles.

## 5.4. Implications for policy design in constrained programmes

Although the model is calibrated to a specific Civil Engineering programme, several policy lessons generalise to other highly structured degrees with strong prerequisite chains.

First, reforms that target *visible inefficiencies*—such as high finals debt or prolonged completion times—can backfire if they ignore underlying structural constraints. Removing regularity and finals without adding support simply moves failure earlier in the trajectory. Institutions should be wary of reforms that promise "swift completion" by tightening progression rules without commensurate investment in teaching, tutoring and remedial infrastructure.

Second, early bottleneck courses require special attention. The Promotion Wall emerges where high friction, centrality in the curriculum graph and strict progression rules coincide. Interventions at these nodes—such as redesigning assessment, embedding supplemental instruction, or adding targeted summer modules—are likely to have disproportionately large effects on cohort survival and equity (Geisinger & Raman, 2013; Paz, 2025a). The safety-net scenario illustrates one such intervention; others could be explored within the same modelling framework.

Third, equity considerations cannot be reduced to overall dropout rates. The model shows that a policy can reduce debt and even maintain similar aggregate completion while substantially worsening outcomes for low-resilience students. Routine monitoring of disaggregated indicators—by resilience proxies, socio-economic status, or first-generation status—is therefore essential. Agent-based simulations can complement this monitoring by stress-testing proposed rules under heterogeneous profiles before real students are exposed to them.

Finally, the findings support a shift from one-off reforms towards *policy portfolios* that combine structural, instructional and psychosocial components. For example, a viable alternative to the pure promotion regime might include: (a) moderated promotion thresholds in first-year bottlenecks; (b) guaranteed remedial capacity focused on low-resilience profiles; and (c) parallel initiatives to foster peer support and belonging in early cohorts. While such portfolios are more complex to implement, they are more likely to move the institution to a better point on the efficiency–equity frontier.

## 5.5. Methodological contributions and broader relevance

Beyond the substantive case, the study contributes methodologically to the emerging use of agent-based models in higher education. First, it demonstrates that ABMs can be **empirically anchored** through calibration to survival curves and archetype distributions derived from real data, addressing concerns that educational simulations often rely on ad hoc assumptions (Grimm et al., 2020; Hamill & Gilbert, 2016). Second, by explicitly representing institutional rules and psychological mechanisms, the model offers a transparent mapping from policy levers to outcomes, in contrast to "black-box" predictive models that treat dropout as a static classification problem.

Third, the Promotion Wall experiment illustrates how ABMs can support **ex ante evaluation** of reforms that would be ethically or practically difficult to trial at scale. The institution analysed here has already experienced a shift towards direct promotion; the model helps disentangle which components of that reform drive observed changes and which counterfactual combinations (e.g., adding safety nets) might mitigate harm. Similar approaches could be applied to other policy questions, such as changing credit loads, reconfiguring prerequisite structures, or introducing adaptive assessment regimes.

Finally, the work highlights the value of integrating **psychological constructs**—stress, belonging, resilience—into structural models of progression. Doing so enables richer interpretations of policy effects and connects institutional analytics to a substantial body of psychological research on motivation and persistence (Martin & Marsh, 2008; Robotham & Julian, 2006). In the context of CAPIRE, this integration paves the way for future studies that combine simulation with qualitative accounts of student experience, closing the loop between computational modelling and lived reality.

In sum, the findings suggest that policy regimes in highly constrained programmes should be evaluated not only by their formal rules but by the trajectories they generate: who is pushed against the wall, who is offered a safety net, and at what psychological cost. The next section turns to the limitations of the present study and outlines avenues for future work.

## 6. LIMITATIONS AND FUTURE WORK

Like any computational experiment, the Promotion Wall study trades realism for tractability. Several limitations should be acknowledged before drawing strong policy conclusions, and they point naturally to extensions of the CAPIRE Intervention Lab.

### 6.1. Scope and calibration of the empirical baseline

The model is calibrated to a *single* Civil Engineering programme in one public university. Although the structural features of this curriculum (long duration, tight prerequisite chains, early mathematical bottlenecks) are common in engineering worldwide (Geisinger & Raman, 2013), the specific parameter values—failure rates, debt patterns, survival curve—are local. External validity is therefore analytic rather than statistical: the goal is to clarify mechanisms rather than to produce generalisable numerical estimates (Epstein, 2006; Railsback & Grimm, 2019).

Moreover, the calibration targets focus on **dropout timing**, not graduation or intermediate outcomes such as credit accumulation or GPA. Completion data are available but were not explicitly fitted, in order to avoid over-constraining a model that already incorporates many micro-processes. It is therefore possible that some parameter combinations consistent with the dropout curve produce unrealistic completion patterns. Future work should extend calibration to a broader set of indicators, for example by matching simulated and empirical joint distributions of time-to-dropout and time-to-degree (Fagiolo et al., 2019).

Finally, the survival analysis underlying the empirical reference curve makes assumptions about censoring and cohort definition (Paz, 2022). Alternative operationalisations—for instance, using shorter inactivity windows or distinguishing temporary stop-outs from permanent leavers—might shift the calibration targets and, consequently, parameter values. A useful extension would subject the model to *calibration under uncertainty*, exploring how conclusions change under different but plausible survival specifications.

**6.2. Simplified behavioural and psychological dynamics**

The model represents students as archetypes with stable traits (ability, resilience, planning style) and two psychological states (stress, belonging). These constructs capture important dimensions of the retention literature (Martin & Marsh, 2008; Respondek et al., 2017), but remain highly stylised. Stress and belonging follow simple linear update rules, and the hazard of dropout is a logistic function of their current levels.

In reality, emotional trajectories can exhibit non-linearities, habituation and path dependence—students may "bounce back" after early failures or, conversely, experience sudden collapses after a critical event (Pekrun & Linnenbrink-García, 2014). Likewise, resilience is treated as an exogenous label assigned at entry, whereas evidence suggests that coping resources are shaped over time by experiences, social support and institutional messages (Cabrera et al., 1993; Walton & Cohen, 2007).

Future work should therefore consider richer behavioural rules, such as:

- adaptive workload choice based on past performance and perceived control;
- peer effects, whereby stress and belonging diffuse through study groups or cohorts;

- explicit modelling of help-seeking and feedback loops between remedial success and resilience.

These additions would increase complexity but could be introduced modularly, with careful sensitivity analysis to avoid "overfitting by story" (Grimm et al., 2020; Squazzoni et al., 2020).

**6.3. Policy space and implementation realism**

The study explores only three stylised policy regimes: historical regularity, pure direct promotion and direct promotion with a fixed-capacity safety net. Real institutions implement more nuanced blends: conditional promotion based on grade bands, varying time-to-live rules for regularity, differentiated support for first-generation students, or adaptive remedial capacity that responds to demand and budgets.

The safety-net scenario, in particular, assumes a constant remedial capacity equal to 30% of active students per year and an ideal allocation mechanism that perfectly observes resilience. In practice, capacity is mediated by teaching resources, funding cycles and scheduling constraints; identifying vulnerable students requires imperfect proxies, such as prior grades or socio-demographic indicators. Modelling such frictions (Herd & Moynihan, 2019) would likely reduce the effectiveness of the safety net and could reveal new equity trade-offs, for example when support is systematically under-allocated to "quietly struggling" students.

A natural next step is to enlarge the policy design space and treat it as an optimisation problem: search over combinations of promotion thresholds, remedial capacity and targeting rules to map out a more complete efficiency–equity frontier. Techniques from reinforcement learning or evolutionary search could be embedded in the ABM for this purpose, while keeping empirical calibration as a hard constraint.

**6.4. Costs, benefits and broader outcomes**

The current model treats remedial capacity as an abstract constraint and does not attach explicit **financial or organisational costs** to policies. Nor does it account for potential benefits beyond dropout reduction, such as improved learning outcomes, employability or student well-being. As a result, the policy comparison remains partly qualitative: Scenario C is clearly more equitable than B, but its cost-effectiveness relative to A cannot be quantified.

Extending the model with basic cost structures—staff time, classroom space, stipends—and simple benefit metrics—credits accumulated, proportion of students reaching key milestones—would enable more policy-relevant analyses. For example, institutions could ask whether reallocating resources from broad but shallow support (e.g., generic tutoring centres) towards targeted, intensive remediation at bottlenecks yields better returns, or how the Promotion Wall interacts with limited scholarship funding windows.

**6.5. Single-cohort focus and systemic feedbacks**

Simulations in this study track a single intake cohort in isolation. In reality, universities operate with **overlapping cohorts**, shared capacity constraints and feedback loops between past and future policies. For instance, high early dropout may free capacity in later years but damage the institution's reputation, affecting future applications; remedial programmes that prove successful may attract additional funding or change staff attitudes. None of these system-level dynamics are represented.

A promising direction is therefore to extend the Intervention Lab into a *multi-cohort, multi-year* model, in which decisions about capacity, staffing and policy are updated based on observed outcomes. Such a model could explore questions about long-term equilibrium: Does the Promotion Wall lead to a smaller but more selective programme? Under what conditions can a safety net become financially self-sustaining through improved completion? These questions connect the micro-dynamics of attrition with meso-level institutional strategy (DeAngelo et al., 2011; Tinto, 2012).

**6.6. Data, transparency and reproducibility**

Finally, while the present implementation follows best-practice guidelines for documenting agent-based models (Grimm et al., 2020; Railsback & Grimm, 2019), it still faces limitations of **data access** and replicability. Administrative micro-data are sensitive, and even anonymised trajectories may not be publicly shareable. Calibration and validation rely on aggregate statistics and survival curves that can be published, but external researchers may not be able to reproduce every detail without access to the raw data.

To mitigate this, future releases of the CAPIRE Intervention Lab will include:

- synthetic datasets generated from the calibrated model, which approximate the joint distributions of key variables without disclosing real students' records;
- open-source code with full configuration files, audit scripts and documentation;
- detailed methodological appendices explaining calibration decisions and alternative parameter sets.

Such transparency is essential if agent-based policy experiments are to inform real decisions rather than remain internal tools (Epstein, 2006; Squazzoni et al., 2020).

In summary, the Promotion Wall experiment should be understood as a *proof of concept*: it shows that progression rules, psychological dynamics and equity considerations can be integrated into a single computational framework grounded in real data. Addressing the limitations outlined above will strengthen both the scientific robustness and the practical usefulness of the CAPIRE Intervention Lab, paving the way for a richer dialogue between institutional analytics, educational research and policy design.

## 7. CONCLUSIONS

This article used a calibrated agent-based model to examine how alternative progression rules reshape student trajectories, psychological experiences and equity patterns in a highly constraining engineering programme. Building on the CAPIRE framework and an empirical survival analysis of 1,343 students, we simulated three policy regimes: a historical regularity + finals regime (Scenario A), a direct-promotion regime (Scenario B), and a direct-promotion regime with a remedial safety net (Scenario C).

The findings support the Promotion Wall hypothesis. When regularity and finals debt are replaced by strict direct promotion in a curriculum with first-year bottleneck courses, attrition becomes sharply front-loaded: dropout concentrates in the first two years, overall dropout rises, and inequities between low- and high-resilience students widen. In contrast, the historical regime produces more gradual "leakage" over time, with moderate dropout but high stocks of unresolved academic debt. Introducing a targeted safety net—capacity-limited remedial courses prioritising vulnerable students—partially dismantles the wall: dropout and equity gaps decrease relative to pure direct promotion, and final stress levels are lowest among the three regimes.

These results highlight three substantive points. First, progression rules are not neutral administrative details; they are structural mechanisms that decide *who* is allowed time to integrate academically and socially, and *when* exclusion happens. Any serious account of dropout and retention must therefore treat institutional rules as causal factors on a par with individual characteristics. Second, apparent improvements in efficiency—such as "clean" transcripts and reduced finals debt—can conceal deeper losses in inclusion and fairness if vulnerable students are disproportionately pushed out early. Efficiency claims that ignore distributional effects are, at best, incomplete. Third, equity-enhancing policies are not cost-free. The safety-net regime improves outcomes only because it injects additional teaching capacity and targets it intentionally at students who would otherwise be filtered out by the Promotion Wall.

Methodologically, the study demonstrates the value of combining a leakage-aware data layer, survival-based calibration and agent-based simulation within a single intervention lab. This integration makes it possible to explore counterfactual policy regimes that are difficult or impossible to test experimentally, while maintaining explicit links to observed data and well-developed theories of stress, belonging and resilience. In doing so, the CAPIRE Intervention Lab complements—but does not replace—traditional statistical analyses: it provides a sandbox in which institutional actors can rehearse potential reforms and inspect not only aggregate outcomes but also their temporal and distributive profiles.

For practitioners and policy-makers in engineering education and other tightly structured programmes, the central message is straightforward. Reforms that tighten progression rules without parallel investment in support structures are likely to build higher walls rather than stronger bridges. If the goal is to reduce time-to-degree *and* widen participation, policy portfolios must combine structural redesign (of curricula and rules), instructional improvement (especially in bottleneck courses) and psychosocial support (including targeted remediation). The Promotion Wall experiment offers one quantitative articulation of how such combinations might work; future extensions of CAPIRE can broaden the policy space and deepen the dialogue between models, evidence and institutional judgment.

## 8. REFERENCES

Abramson, L. Y., Seligman, M. E. P., & Teasdale, J. D. (1978). Learned helplessness in humans: Critique and reformulation. *Journal of Abnormal Psychology, 87*(1), 49–74.

Aljohani, O. (2016). A comprehensive review of the major studies and theoretical models of student retention in higher education. *Higher Education Studies, 6*(2), 1–18. https://doi.org/10.5539/hes.v6n2p1

Bandura, A. (1997). *Self-efficacy: The exercise of control*. W. H. Freeman.

Bean, J. P., & Metzner, B. S. (1985). A conceptual model of nontraditional undergraduate student attrition. *Review of Educational Research, 55*(4), 485–540.

Bensimon, E. M. (2007). The underestimated significance of practitioner knowledge in the scholarship on student success. *The Review of Higher Education, 30*(4), 441–469.

Bound, J., Lovenheim, M. F., & Turner, S. (2010). Why have college completion rates declined? An analysis of changing student preparation and collegiate resources. *American Economic Journal: Applied Economics, 2*(3), 129–157.

Boudon, R. (1974). *Education, opportunity, and social inequality: Changing prospects in Western society*. Wiley.

Cabrera, A. F., Nora, A., & Castañeda, M. B. (1993). College persistence: Structural equations modeling test of an integrated model of student retention. *Journal of Higher Education, 64*(2), 123–139.

DeAngelo, L., Franke, R., Hurtado, S., Pryor, J., & Tran, S. (2011). *Completing college: Assessing graduation rates at four-year institutions*. Higher Education Research Institute, UCLA.

DesJardins, S. L., Ahlburg, D. A., & McCall, B. P. (2002). A temporal investigation of factors related to timely degree completion. *Journal of Higher Education, 73*(5), 555–581.

DiPrete, T. A., & Eirich, G. M. (2006). Cumulative advantage as a mechanism for inequality: A review of theoretical and empirical developments. *Annual Review of Sociology, 32*, 271–297.

Dowd, A. C., & Bensimon, E. M. (2015). *Engaging the "race question": Accountability and equity in U.S. higher education*. Teachers College Press.

Epstein, J. M. (2006). *Generative social science: Studies in agent-based computational modeling*. Princeton University Press.

Ezcurra, A. M. (2019). *Desigualdades en la educación superior: Acceso, permanencia y graduación en la universidad*. Miño y Dávila.

Fagiolo, G., Guerini, M., Lamperti, F., Moneta, A., & Roventini, A. (2019). Validation of agent-based models in economics and finance. In C. Hommes & B. LeBaron (Eds.), *Handbook of computational economics* (Vol. 4, pp. 613–675). Elsevier.

Geisinger, B. N., & Raman, D. R. (2013). Why they leave: Understanding student attrition from engineering majors. *International Journal of Engineering Education, 29*(4), 914–925.

Grimm, V., Railsback, S. F., Vincenot, C. E., Berger, U., Gallagher, C., DeAngelis, D. L., Edmonds, B., Ge, J., Giske, J., Groeneveld, J., Johnston, A. S. A., Milles, A., Nabe-Nielsen, J., Polhill, J. G., Radchuk, V., Rohwäder, M. S., Stillman, R. A., Thiele, J. C., & Ayllón, D. (2020). The ODD protocol for describing agent-based and other simulation models: A second update to improve clarity, replication, and structural realism. *Journal of Artificial Societies and Social Simulation, 23*(2), Article 7. https://doi.org/10.18564/jasss.4259

Hamill, L., & Gilbert, N. (2016). *Agent-based modelling in economics*. Wiley.

Herd, P., & Moynihan, D. P. (2019). *Administrative burden: Policymaking by other means*. Russell Sage Foundation.

Kahu, E. R., & Nelson, K. (2018). Student engagement in the educational interface: Understanding the mechanisms of student success. *Higher Education Research & Development, 37*(1), 58–71.

Kaplan, E. L., & Meier, P. (1958). Nonparametric estimation from incomplete observations. *Journal of the American Statistical Association, 53*(282), 457–481.

Martin, A. J., & Marsh, H. W. (2008). Academic resilience and academic buoyancy: Multidimensional and hierarchical conceptual framing of causes, correlates and cognate constructs. *Oxford Review of Education, 34*(3), 353–370.

Nora, A. (2004). The role of habitus and cultural capital in choosing a college, transitioning from high school to higher education, and persisting in college among minority and nonminority students. *Journal of Hispanic Higher Education, 3*(2), 180–208.

OECD. (2019). *Education at a glance 2019: OECD indicators*. OECD Publishing.

Paz, H. R. (2022). Determinación del tiempo medio de deserción y de los factores que facilitan o retrasan la deserción estudiantil en una carrera de Ingeniería Civil. *Revista Latinoamericana de Políticas y Administración de la Educación, 17*, 99–116.

Paz, H. R. (2025a). An agent-based simulation of regularity-driven student attrition: How institutional time-to-live constraints create a dropout trap in higher education. *arXiv*. https://doi.org/10.48550/arXiv.2511.16243

Paz, H. R. (2025b). A leakage-aware data layer for student analytics: The CAPIRE framework for multilevel trajectory modelling. *arXiv preprint arXiv:2511.11866*. https://doi.org/10.48550/arXiv.2511.11866

Pekrun, R., & Linnenbrink-García, L. (2014). *International handbook of emotions in education*. Routledge.

Railsback, S. F., & Grimm, V. (2019). *Agent-based and individual-based modeling: A practical introduction* (2nd ed.). Princeton University Press.

Respondek, L., Seufert, T., Stupnisky, R. H., & Nett, U. E. (2017). Perceived academic control and academic emotions predict undergraduate university student success: Examining effects on dropout intention and achievement. *Frontiers in Psychology, 8*, 243.

Robotham, D., & Julian, C. (2006). Stress and the higher education student: A critical review of the literature. *Journal of Further and Higher Education, 30*(2), 107–117.

Squazzoni, F., Ghorbani, A., Bravo, G., Jager, W., Keesman, N., Davidsen, P. I., Holtz, G., Wijermans, N., Giardini, F., & Edmonds, B. (2020). Computational models that matter during a global pandemic outbreak: A call to action. *Journal of Artificial Societies and Social Simulation, 23*(2), Article 10.

Tierney, W. G. (1992). An anthropological analysis of student participation in college. *Journal of Higher Education, 63*(6), 603–618.

Tilly, C. (1998). *Durable inequality*. University of California Press.

Tinto, V. (1975). Dropout from higher education: A theoretical synthesis of recent research. *Review of Educational Research, 45*(1), 89–125.

Tinto, V. (1993). *Leaving college: Rethinking the causes and cures of student attrition* (2nd ed.). University of Chicago Press.

Tinto, V. (2012). *Completing college: Rethinking institutional action*. University of Chicago Press.

UNESCO-IESALC. (2020). *COVID-19 and higher education: Today and tomorrow*. UNESCO International Institute for Higher Education in Latin America and the Caribbean.

Walton, G. M., & Cohen, G. L. (2007). A question of belonging: Race, social fit, and achievement. *Journal of Personality and Social Psychology, 92*(1), 82–96.

Witham, K., & Bensimon, E. M. (2012). Creating a culture of inquiry around equity and student success. In S. D. Museus & U. M. Jayakumar (Eds.), *Creating campus cultures: Fostering success among racially diverse student populations* (pp. 115–134). Routledge.

Yosso, T. J., Smith, W. A., Ceja, M., & Solórzano, D. G. (2009). Critical race theory, racial microaggressions, and campus racial climate for Latina/o undergraduates. *Harvard Educational Review, 79*(4), 659–691.